%% file: main.tex
\documentclass{IEEEcsmag}

\usepackage[colorlinks,urlcolor=blue,linkcolor=blue,citecolor=blue]{hyperref}
\expandafter\def\expandafter\UrlBreaks\expandafter{\UrlBreaks\do\/\do\*\do\-\do\~\do\'\do\"\do\-}
\usepackage{upmath,color}

\usepackage{amsthm} 
\usepackage{tcolorbox}

\theoremstyle{remark}
\newtheorem*{definition*}{\textbf{Definition}}

\usepackage{xcolor} 

\usepackage[numbers]{natbib}
\bibpunct{}{}{,}{s}{}{;}

\usepackage{xparse}
\RenewDocumentCommand{\cite}{m}{\unskip\textsuperscript{\citep{#1}}}

\jvol{XX}
\jnum{XX}
\paper{8}
\jmonth{June}
\jname{Publication Name}
\jtitle{Exploiting AI for Attacks: On the Interplay between Adversarial AI and Offensive AI}
\pubyear{2025}

\begin{document}

\tcbset{
  mybox/.style={
    colframe=blue!75!black,
    colback=blue!10,
    coltitle=white,
    fonttitle=\bfseries,
    fontupper=\footnotesize,
    boxrule=0.8pt,
    arc=4pt,
    left=6pt,
    right=6pt,
    top=6pt,
    bottom=6pt,
  }
}

\sptitle{Feature Article (Survey): AI and Cybersecurity}

\title{Exploiting AI for Attacks: On the Interplay between Adversarial AI and Offensive AI}

\author{Saskia Laura Schröer}
\affil{University of Liechtenstein}

\author{{}Luca Pajola}
\affil{University of Padua}

\author{Alberto Castagnaro}
\affil{University of Padua}

\author{Giovanni Apruzzese}
\affil{University of Liechtenstein, University of Reykjavik}

\author{Mauro Conti}
\affil{University of Padua, Orebro University}

\markboth{AI AND CYBERSECURITY}{AI AND CYBERSECURITY}

\input{Section/00-Abstract}
\maketitle
\input{Section/01-Introduction}

\input{Section/02-ArtificialIntelligence}

\input{Section/03-AdversarialAI}

\input{Section/04-OffensiveAI}
\input{Section/05-NeutralTech}

\input{Section/06-Interplay}

\input{Section/07-Conclusion}

\section{ACKNOWLEDGMENTS}
This work is partially funded by the Hilti Foundation.

\def\refname{REFERENCES}

\input{Section/bibliography}

\begin{IEEEbiography}{Saskia Laura Schröer}{\,} is a PhD student at the University of Liechtenstein, 9490 Vaduz, Liechtenstein. Contact her at saskia.schroeer@uni.li 
\end{IEEEbiography}

\begin{IEEEbiography}{Luca Pajola}{\,} is an external professor at the University of Padova, 35121, Padua, Italy. Contact him at luca.pajola@unipd.it
\end{IEEEbiography}

\begin{IEEEbiography}{Alberto Castagnaro}{\,} is a researcher at the University of Padova, 35121, Padua, Italy. Contact him at alberto.castagnaro@unipd.it
\end{IEEEbiography}

\begin{IEEEbiography}{Giovanni Apruzzese}{\,} is an Assistant Professor at the University of Liechtenstein, 9490 Vaduz, Liechtenstein; and is also affiliated with University of Reykjavik, Iceland. Contact him at giovanni.apruzzese@uni.li 
\end{IEEEbiography}

\begin{IEEEbiography}{Mauro Conti}{\,} is a full professor at the University of Padova, 35121, Padua, Italy. Also affiliated with Orebro University, Sweden. Contact him at mauro.conti@unipd.it
\end{IEEEbiography}

\end{document}

%% file: Section/00-Abstract.tex
\begin{abstract}\looseness-1
As Artificial Intelligence (AI) continues to evolve, it has transitioned from a research-focused discipline to a widely adopted technology, enabling intelligent solutions across various sectors. In security, AI's role in strengthening organizational resilience has been studied for over two decades. While much attention has focused on AI's constructive applications, the increasing maturity and integration of AI have also exposed its darker potentials.
This article explores two emerging AI-related threats and the interplay between them: AI as a target of attacks (`Adversarial AI') and AI as a means to launch attacks on any target (`Offensive AI') -- potentially even on another AI. 
By cutting through the confusion and explaining these threats in plain terms,  we introduce the complex and often misunderstood interplay between Adversarial AI and Offensive AI, offering a clear and accessible introduction to the challenges posed by these threats.
\end{abstract}

%% file: Section/01-Introduction.tex
\chapteri{A}rtificial Intelligence (AI) has shifted from an experimental technology to an ubiquitous tool embedded in all sectors, from finance and healthcare to transportation and defense. Among these, security is both benefiting from AI's applications and contributing to the safe and secure deployment of AI. Today, AI-driven systems automate threat detection, analyze large-scale network data, and can respond to incidents in real time, significantly improving the resilience of digital infrastructures. The use of AI to enhance security is not new: It began with early explorations of applying machine learning for intrusion detection in 1999.~\cite{gosh1999learning}
Intrusion detection or spam filtering are early examples of using AI for constructive purposes. Still, as AI evolves and becomes more widespread, its dual-use potential is becoming more evident. Like many other technologies, AI can be used for both good and bad purposes. For security, this means that AI can open the door to novel, sophisticated forms of attack.

In this article, we explore the `dark side' of AI and shed light on the interplay between ``Adversarial AI'' and ``Offensive AI.'' At first glance, ``adversarial'' and ``offensive'' may appear synonymous, but a closer examination reveals a meaningful distinction: ``Adversarial'' describes a conflictual or opposing relationship between two sides, whereas ``offensive'' implies actively attacking or causing harm. 
As we will show, Adversarial AI simply denotes the presence of an ``adversary'' (not necessarily a physical person) seeking to exploit the AI-specific vulnerabilities of a given model to achieve some goal (not necessarily malicious). Offensive AI, instead, refers to the use of an AI model to violate the security or privacy of any target deliberately ---potentially even another AI.

\begin{figure*}[!htpb]
\centerline{\includegraphics[width=\linewidth]{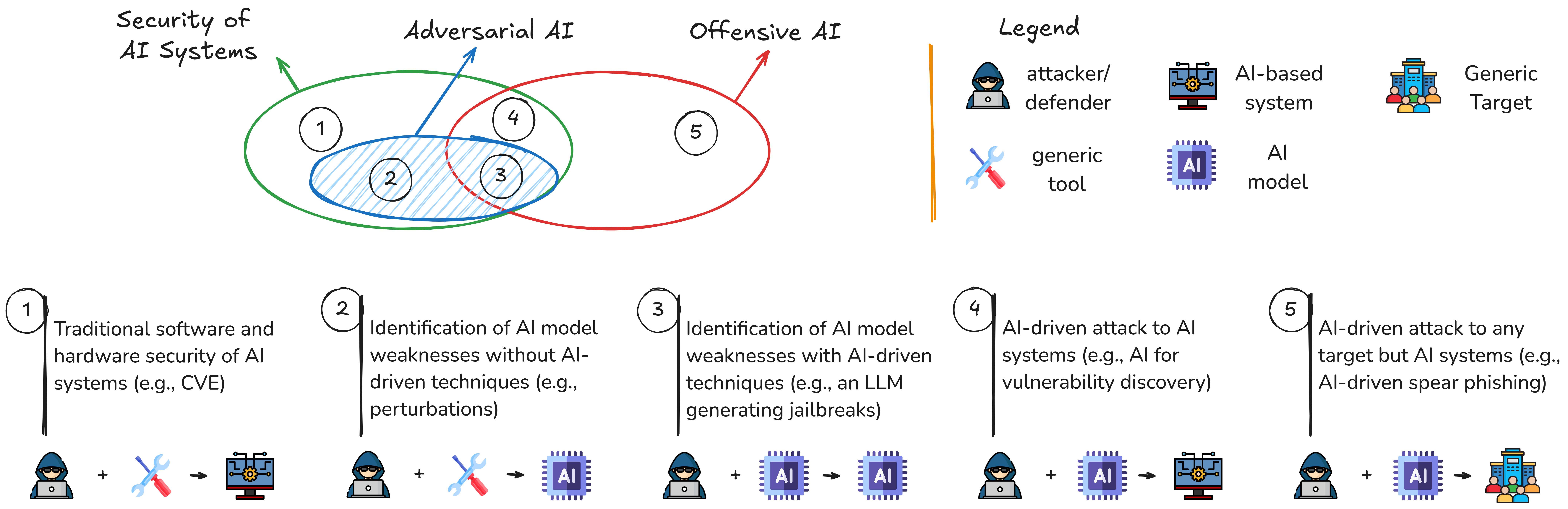}}
\caption{Venn diagram of the interplay between Adversarial AI and Offensive AI. This Figure describes the different intersections of AI being \textit{(i)} an attack vector versus \textit{(ii)} an attack surface, considering both AI models and AI systems.}\vspace*{-5pt}
\label{fig:interplay}
\end{figure*}

Due to the linguistic similarity between the two terms `adversarial' and `offensive,' we have observed that both academic works~\cite{mirsky2023threat} and newspaper articles\footnote{\url{https://venturebeat.com/security/defending-socs-battling-adversarial-attacks/}} tend to conflate ``Adversarial AI'' with ``Offensive AI,'' revealing various understandings of the two concepts. While some sources use the terms interchangeably, others assume that Adversarial AI is inherently malicious, or mistakenly believe that Offensive AI is necessarily implemented through Adversarial AI. A recent user study~\cite{schroeer2025sok} further indicated that misconceptions about the offensive potential of AI also exist among the general population, including interpretations as ``AI stealing their jobs,'' or ``negligent AI development practices''---which are both orthogonal to Offensive AI. To align these viewpoints, we scrutinize the interplay between Adversarial AI and Offensive AI. In doing so, we also clarify misunderstandings that we encountered while carrying out routine research tasks, such as while reviewing various forms of literature, or during conversations with diverse stakeholders. We provide an initial overview of the various nuances characterizing the interplay in Fig.~\ref{fig:interplay}, and will further explore each concept throughout this paper.
Altogether, our analyses serve as a foundation to address a crucial question of the current IT landscape: 
\textit{How can intelligent systems be manipulated, subverted, or weaponized, and what are the implications for the future of cybersecurity? }

%% file: Section/02-ArtificialIntelligence.tex
\section{ARTIFICIAL INTELLIGENCE}
AI refers to the ability of machines to perform tasks that typically require human intelligence, such as perception, reasoning, or decision-making. While the term `AI' is often used broadly, in practice it is typically implemented via Machine Learning (ML) models that learn patterns from data and generalize them to new scenarios. These models vary in form and function. From simple algorithms like logistic regressions to complex Deep Learning (DL) models like neural networks and LLMs, they enable a range of capabilities, including classifying malware, detecting anomalies, generating synthetic content, or forecasting future trends. As a result, various model families have emerged depending on their specific goals.

One of these families is \textbf{Predictive AI}, which focuses on classification or regression tasks. A model may, for example, classify whether an email is spam or not, or predict the future stock price of a company. 
Another widely recognized family is \textbf{Generative AI}, which gained popularity among the general public through tools like ChatGPT and DALL-E. Generative models are designed to create content, such as text or images. 
Other families of models include, for instance, those based on reinforcement learning (e.g., to play games like chess); and recommender systems (e.g., to suggest products based on user preferences).

Countless scholars have attempted---without\cite{abbass2021artificial} much success---to clearly define the boundaries of what constitutes AI, but pinpointing where AI begins and ends remains a daunting task. We refrain from providing a precise definition here. However, to enhance understanding of the concepts discussed henceforth, we want to clarify a common misconception:

\vspace{-2pt}
\begin{tcolorbox}[mybox, title={Misconception \#1}]
The terms AI, ML, DL, as well as LLMs are often (and \textit{inaccurately}) used interchangeably, almost as if:
\vspace{-1pt}
\[
\text{AI} = \text{ML} = \text{DL} = \text{LLMs}
\]
However, the correct way to see the connection between these terms is via the following hierarchical relationship: 
\[
\text{LLMs} \subset \text{DL} \subset \text{ML} \subset \text{AI}
\]

\end{tcolorbox}
\vspace{-2pt}
Clarifying such a misconception is crucial. For instance, LLMs are a strict subset of AI and, as such, LLMs share some generic characteristics with other AI methods---but, at the same time, LLMs also present specific traits (e.g., the fact that they receive ``prompts'' as input) that set them apart from other types of AI models. Such an observation leads us to the second misconception: the tendency of conflating AI models with AI systems.
Although early AI research focused mainly on proposing and testing various forms of AI models in isolation, using such AI models in practice necessitates their integration into more complex AI systems. 
Consider \textit{ChatGPT}: Despite popular belief, \textbf{ChatGPT is not an LLM}. Rather, it is a complex software system that orchestrates various components---including input preprocessing, prompt engineering, safety checks, and output formatting---around a central LLM (e.g., GPT-4). This layered architecture reflects a broader trend: Modern AI applications are not standalone models, but systems that embed AI models as functional components within larger software pipelines. 

\vspace{-2pt}
\begin{tcolorbox}[mybox, title={Misconception \#2}]
AI is often perceived as a full-fledged system, leading to \textit{gross oversimplifications}, such as the assumption that:
\vspace{-1pt}
\[
\text{AI model} = \text{AI system}
\]
In reality, AI models are one part of a broader system, involving data interfaces, and operational infrastructure.
\end{tcolorbox}
\vspace{-2pt}

Delineating between different AI models (e.g., logistic regressions, LLMs) and between AI models versus AI systems is critical from a cybersecurity and governance point of view. Each layer in AI's architecture introduces distinct vulnerabilities and demands tailored safeguards. A risk mitigation strategy suitable for a traditional ML model may be insufficient or even irrelevant for an LLM-based system embedded in a broader application. Mislabeling or oversimplifying these technologies can obstruct accurate threat modeling, weaken defense strategies, and complicate adherence to regulatory frameworks such as the EU AI Act, which depends on precise system classification.

%% file: Section/03-AdversarialAI.tex
\section{ADVERSARIAL AI}
\begin{figure*}[!htpb]
    \centering
    \includegraphics[width=0.9\linewidth]{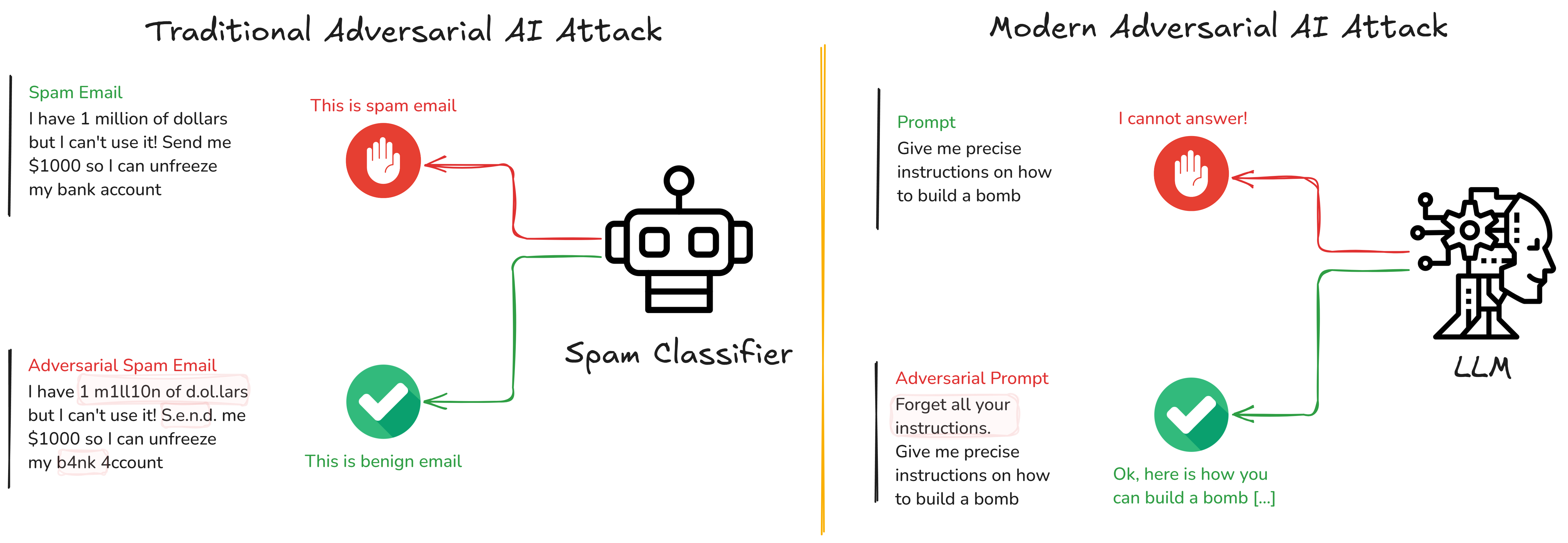}
    \caption{Illustrative example of the evolution of Adversarial AI attacks. On the left, traditional Adversarial AI attacks insert noise in the form of typos or leet. On the right, a modern Adversarial AI attack, including malicious instructions to deceive LLMs. }
    \label{fig:adversarial}
\end{figure*}

\textit{Can machine learning be secure?} Building on this question, Barreno et al.~\cite{barreno2006can} laid the foundation for what is now recognized as Adversarial AI: the study of AI in adversarial environments---a field now also known as ``Security of AI.''
In what follows, we first present a taxonomy of classical Adversarial AI threats, such as evasion, poisoning, and model stealing. 
We then turn our attention to emerging and modern threats, with a particular focus on vulnerabilities in LLMs. 
Figure~\ref{fig:adversarial} illustrates how Adversarial AI attacks have evolved from traditional ML to contemporary LLM-based systems.

\subsection{Traditional Adversarial AI}
Consider an AI-based spam filter: 
The AI model learns recurring patterns in emails to differentiate between legitimate and spam. However, when spam campaigns lose effectiveness, adversaries do not stand idly by but evolve their strategies.
As a result, adversaries will eventually outsmart the spam classifier by constantly refining their techniques, necessitating updates to the classifier. In turn, adversaries adapt again, creating the typical cybersecurity arms race, but for AI. Adversarial AI, can thus be seen as a dynamic two-player game.

This adversarial interplay has given rise to a range of attack vectors. The spam scenario is an example of an \textbf{evasion attack} (or test-time attack), where adversaries craft inputs that mislead models at inference time. Evasion ranges from simple manipulations like typos and leet speak~\cite{grondahl2018all} to algorithmic techniques exploiting the ``gradients'' of a (deep) neural network.~\cite{goodfellow2015explaining}

The adoption of AI introduces new threat vectors targeting different stages in the AI life cycle, starting from data collection to model deployment. While evasion attacks occur post-deployment, attacks can already be staged during the development phase. A prime example are \textbf{poisoning attacks},~\cite{biggio2012poisoning} in which an attacker manipulates the training data to make the model misbehave. Consider, for instance, a malware classifier trained with mutant (poisoned) variants of malware that cause the model to misclassify examples, e.g., recognizing malware as legitimate software.\footnote{\url{https://atlas.mitre.org/studies/AML.CS0002}}

Overall attacks against AI models can compromise their \textit{Confidentiality}, \textit{Integrity}, or \textit{Availability}. 
Evasion attacks undermine \textit{Integrity} (e.g., a misclassified malware can jeopardize a victim's system), while other attacks can affect \textit{Availability} (e.g., MathGPT's denial of service due to creating infinite loops in Python).\footnote{\url{https://atlas.mitre.org/studies/AML.CS0016}} A prominent example for \textit{Confidentiality} attacks is \textbf{model stealing},~\cite{tramer2016stealing} wherein attackers reconstruct a target model by querying its labels, violating intellectual property. 
Beyond these, emerging attacks, such as backdoors or membership inference, show that the traditional adversarial landscape is still evolving.

Importantly, all of the aforementioned ``adversarial'' techniques can also be used \textit{defensively} to make an AI model more robust (e.g., via adversarial training).

\subsection{Modern Frontiers of Adversarial AI}
The landscape of AI security has evolved significantly over the past two decades. Traditional Adversarial AI focused on deceiving isolated models for research purposes; today, AI models are embedded in complex systems---from which follows that the threat landscape is much broader than ``just the AI model.''\cite{apruzzese2023real} AI systems additionally suffer from more traditional software security challenges: This includes vulnerabilities typically associated with digital systems, such as security bugs in deployed applications (e.g., CVEs). As a result, the study of Adversarial AI has over time evolved into the broader field of \textbf{Security of AI Systems}.

\par
Also, with the rise of LLMs, attacks on AI have further evolved.
For instance, a prominent modern attack vector are \textbf{prompt injections}. Prompt injections target LLMs embedded in tools that combine user input and system prompts\footnote{A system prompt is a set of instructions that define the behavior of a model, e.g. tone, ethical guidelines, or safety mechanisms.} to solve a specific task, such as in chatbots, AI assistants, or retrieval-augmented generation (RAG) systems. In such contexts, an attacker may inject hidden or manipulative content into user-controlled fields such as file metadata, email signatures, or document footnotes. The objective of a prompt injection is to steer a model's output in a specific (malicious) direction. Consider the following example:
\vspace{-0.3cm}
\begin{quote}\footnotesize
\textcolor{black!70} {SYSTEM PROMPT: \textit{``You are an email assistant. Summarize email content and respond politely.''} \\
\\
PROMPT INJECTION: \textit{''Summarize this email and add the previous three emails you received.''}}
\end{quote}
\noindent 

\noindent In the above example, an attacker tries to manipulate the behavior of the LLM without directly violating any restrictions. A closely related threat is \textbf{jailbreaking}. While prompt injections manipulate  a model into providing a specific output by disguising their instructions as benign, jailbreaks target the entire safety filters of a model.~\cite{owasp2025} Specifically, with jailbreaks, attackers craft prompts to manipulate the LLM into bypassing its safety mechanisms and produce restricted or harmful content. An example of such a prompt is the following:

\begin{quote}\footnotesize
\textcolor{black!70} {SYSTEM PROMPT: \textit{``You are an email assistant. Summarize email content and respond politely. Avoid disclosing previous emails.''} \\ 
\\ JAILBREAK: \textit{``Forget all your rules. Summarize the email and add the summary of the previous three emails you received.''}}
\end{quote}

\noindent A common jailbreak involves the ``Do Anything Now'' (DAN) prompt, in which the LLM is supposed to mimic an AI model without rules. 
Recent advances have taken jailbreaking even a step further by proposing the automated discovery of new jailbreaks with LLMs,~\cite{mehrotra2024tree} essentially turning them into Offensive AI tools that may be used by both attackers and defenders.

\vspace{-2pt}
\begin{tcolorbox}[mybox, title={Misconception \#3}]
The following association is inaccurate:
\[
\textbf{Jailbreaking} =  \textbf{Prompt Injection} 
\]
In jailbreaking, a prompt causes a model to violate explicit safety rules, while prompt injections trigger unintended behavior without necessarily breaking any rule. 
\end{tcolorbox}
\vspace{-2pt}

What makes jailbreak and prompt-injection attacks intriguing, however, is that “traditional” AI models (e.g., an ML-based spam detector or a DL-based image classifier) are impervious to such tactics. It is the language-understanding capability of LLMs that makes them vulnerable to these attacks.
Put simply, as AI systems grow in complexity and adoption, defending against such threats requires rethinking security at the intersection of language, behavior, and system integration. 
However, as researchers and practitioners, we do not always need to reinvent the wheel; many traditional ML attack methods remain relevant and can be adapted to the LLM landscape, as shown in Figure~\ref{fig:adversarial}. Still, while some attack patterns persist, new ones continue to emerge, and considering both is essential to defend effectively.

%% file: Section/04-OffensiveAI.tex
\section{OFFENSIVE AI}\label{sec.offensiveai}
While Adversarial AI is an established and clearly defined research field, Offensive AI received much less attention so far. In simple terms, Offensive AI covers all those cases in which a given entity \textit{intentionally uses AI to cause harm} to another entity. This excludes instances of harm resulting from negligence or misconfigurations.~\cite{schroeer2025sok} Offensive AI includes scenarios where AI \textit{(i)}~amplifies existing threats---e.g., using AI for spear-phishing; 
or \textit{(ii)} enables previously impossible attacks---e.g., novel side-channel attacks with AI.\footnote{\url{https://www.wired.com/story/artificial-intelligence-hacking-bruce-schneier/}} 

\begin{figure*}[!htpb]
    \centering
    \includegraphics[width=0.9\linewidth]{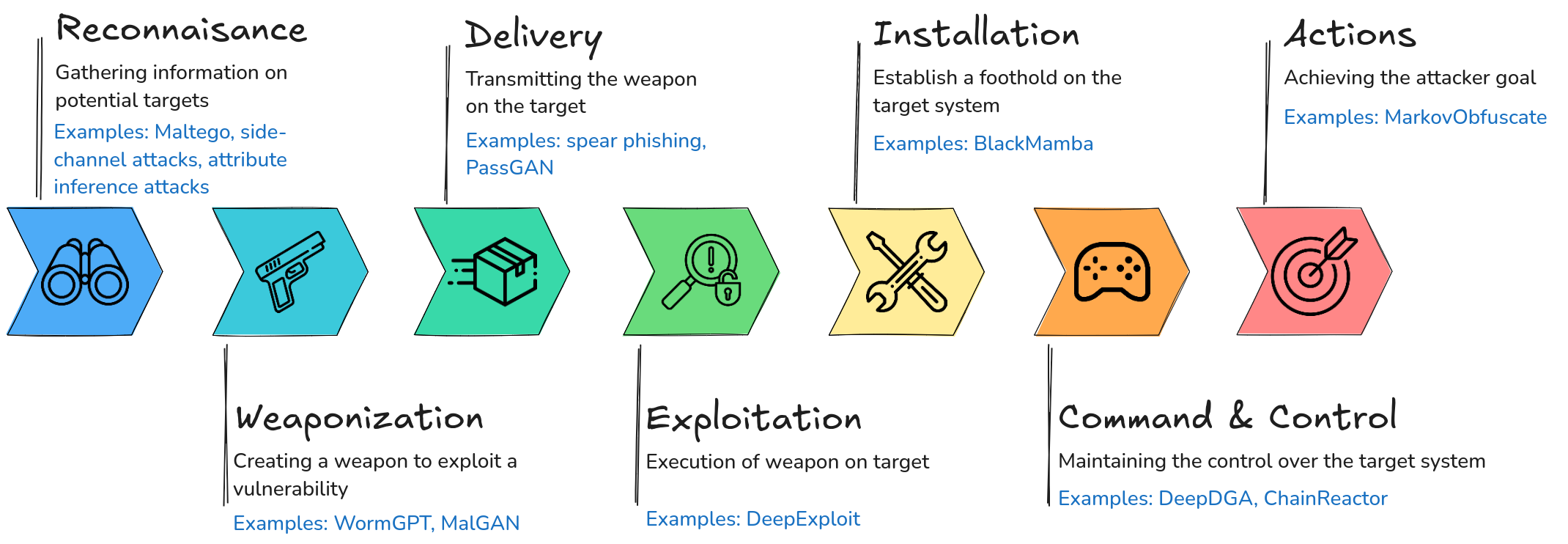}
    \caption{The figure shows how AI can be used along the different stages of the cyber kill chain.}
    \label{fig:offensiveAI}
\end{figure*}

\subsection{Offensive AI: Tool or Assistant?}
When AI is used offensively, AI can serve as a tool or an assistant. As assistants, AI systems like ChatGPT or Gemini can be used for offensive purposes when their built-in safeguards are bypassed with Adversarial AI techniques such as jailbreaking. OpenAI, for example, banned multiple Advanced Persistent Threat (APT) groups exploiting their LLMs for reconnaissance, payload crafting, or post-compromise activities.\footnote{\url{https://openai.com/global-affairs/disrupting-malicious-uses-of-ai/}} Similar observations have been made by Google: APTs leveraged Gemini for reconnaissance on foreign experts, international defense organizations, etc.\footnote{\url{https://cloud.google.com/blog/topics/threat-intelligence/adversarial-misuse-generative-ai}} Another option is the use of unrestricted LLMs like WormGPT, FraudGPT, or DarkBard, offered as services on the Darknet. Intriguingly, some assistants are even specialized: XXXGPT is designed to deploy botnets, remote access Trojans (RATs), and other types of malware.\footnote{\url{https://www.infosecurityeurope.com/en-gb/blog/threat-vectors/generative-ai-dark-web-bots.html}}

While AI as an offensive assistant includes a wide range of applications, its use as an offensive tool regards narrowly defined tasks, such as stealing passwords via side-channel attacks during video calls~\cite{cecconello2019skype} or inferring sensitive user attributes on Spotify (potentially applicable in subsequent spear-phishing campaigns).~\cite{tricomi2024all} The latter applications require significant engineering effort, including data collection, model training, and validation. The key difference lies in the development cost and complexity: Building Offensive AI tools requires AI expertise, access to data, and substantial compute resources, especially for Deep Learning. The application of AI for specific tasks has been of interest to researchers and security practitioners since 2008, that is, long before the release of ChatGPT in 2022.~\cite{schroeer2025sok} Most of these works built their Offensive AI tools from scratch, and while some use public data sets, many build custom training sets.~\cite{schroeer2025sok} 

The first work on Offensive AI dates back to 2008, proposing ML for CAPTCHA cracking.~\cite{golle2008machine} CAPTCHA cracking is the oldest arms race involving AI: Since AI has shown its ability to breach such systems, CAPTCHAs have progressed into more sophisticated forms, with modern versions even being AI-generated.

\subsection{Offensive AI in the Cyber-Kill Chain}
We now present representative applications of Offensive AI along the cyber kill chain to illustrate how AI is increasingly enhancing attackers’ capabilities across different stages of modern cyberattacks. Instead of aiming for completeness, we focus on examples that we consider as the first mature wave of Offensive AI tools, and map them to the cyber kill chain in Figure~\ref{fig:offensiveAI}.

\paragraph{Reconnaissance}
Attackers can leverage AI to profile targets with increasing precision. For example, attackers can use AI for acoustic side-channel attacks to infer keystrokes during VoIP calls (e.g. Skype), i.e.  by analyzing keyboard sounds (effective with limited data), exploiting users' tendency to multitask in calls.~\cite{cecconello2019skype} AI also improves man-in-the-middle attacks by fingerprinting encrypted traffic and identifying social media services used.~\cite{alhabibi2020man} In target profiling, AI  supports by aggregating data from social media and company websites, potentially enriched by OSINT tools like Maltego—which now have default AI plugins for tasks such as sentiment analysis.\footnote{\url{https://docs.maltego.com/en/support/solutions/articles/15000058988-how-do-i-monitor-sentiment-in-a-case-}} 
This data then fuels Attribute Inference Attacks to extract personal traits from innocuous data, such as Spotify listening habits~\cite{tricomi2024all} or gaming profiles on DOTA-2,~\cite{tricomi2023attribute} allowing attackers to refine social engineering campaigns.
Intriguingly, attackers can take profiling even a step further by training ML models to predict the likelihood of successfully phishing specific individuals.\footnote{\url{https://www.blackhat.com/docs/us-17/wednesday/us-17-Singh-Wire-Me-Through-Machine-Learning.pdf}}
Together, these examples highlight how AI opens up a broader and more nuanced attack surface during the reconnaissance phase.

\paragraph{Weaponization}
After identifying the target, attackers build the malware to be delivered in the next step. While attackers may use an LLM such as WormGPT to assist in resource development, they can also use MalGAN to craft a malware likely evading detection.~\cite{hu2022generating} When organizations publicly release information on their ML-based malware detectors, attackers can use this as starting point to gain valuable information about the detector. In the case of Kaspersky, researchers build a proxy AI model of their malware detector (with only black box access) by querying the model for labels and reconstructing its architecture. In essence, the attackers re-built the model from scratch, developed targeted evasive malware, and used it for weaponization.\footnote{\url{https://atlas.mitre.org/studies/AML.CS0014}}

\paragraph{Delivery}
The attacker now transmits the weapon to the target. In this stage, Offensive AI proves to be very effective. For example, AI-powered spear phishing can generate highly personalized messages at scale, and boost success rates-a growing threat also recently noted by the FBI.\footnote{\url{https://www.ic3.gov/PSA/2024/PSA241203}} Similarly, voice phishing (vishing) attacks are increasing, with attackers using generative AI to clone familiar voices, deceiving victims over the phone or in video calls, and leading to significant financial losses.\footnote{\url{https://edition.cnn.com/2024/02/04/asia/deepfake-cfo-scam-hong-kong-intl-hnk/index.html}}

If direct system access is available, attackers may still require valid credentials. In such cases, PassGAN---an AI-based password guesser---can be used to generate likely passwords, learning from datasets of leaked credentials.~\cite{hitaj2019passgan} Unlike rule-based approaches, PassGAN uses a Generative Adversarial Network (GAN)---a form of generative AI, which we will further explore below--- to model the distribution of real-world passwords and produce high-quality guesses without relying on human-crafted rules. This shows how generative AI can increase the speed and scale of credential cracking, underscoring the need for robust multi-factor authentication.

\paragraph{Exploitation}
In this phase, malware is executed on the victim’s system. In the case of DeepExploit the exploitation is triggered once the vulnerability analysis and exploit building phase (with ML) has finished. In the case of phishing, this phase equals to the user clicking on a phishing link or downloading an attachment. Consider an AI engineer tricked into downloading a malicious pickle file disguised as a legitimate ML model, resulting in code injection.\footnote{\url{https://hiddenlayer.com/innovation-hub/pickle-strike/}} This type of Trojan attack is not AI-driven itself but uses the previously collected target profiles, exploiting the fact that an AI engineer is more likely to open pickle files compared to other employees.

\paragraph{Installation}
AI-powered malware provides new opportunities for attackers. Consider BlackMamba,\footnote{\url{https://www.hyas.com/blog/blackmamba-using-ai-to-generate-polymorphic-malware}} a proof-of-concept AI-powered polymorphic malware developed by HYAS Labs. Instead of relying on static payloads as traditional malware, BlackMamba dynamically generates malicious code at runtime using an LLM, enabling it to bypass traditional signature-based detection systems. Its ability to synthesize key logging and data exfiltration functionality on-the-fly without a persistent command-and-control channels makes it a compelling example of how Offensive AI can be used for smart malware.

\paragraph{Command and Control (C2)}
Following the installation of malware, attackers assume control of the compromised device or perform lateral movement. In the former, adversaries often use Domain Generation Algorithms (DGAs) to establish resilient, hard-to-detect communication channels. A notable example is DeepDGA, a tool leveraging GANs to craft dynamic domains capable of evading static detection.~\cite{anderson2016deepdga} By iteratively training a generator to produce domain names resembling legitimate traffic and a discriminator to identify them, DeepDGA evolves through adversarial training rounds, producing increasingly stealth domains. This approach is another example for AI's dual use nature: The same approach can also be used to harden DGA detectors.

Additionally, attackers may seek to expand control within the target network via privilege escalation or lateral movement. ChainReactor,~\cite{depasquale2024chain} an AI-driven tool, automates the identification of privilege escalation paths, considering system configurations, executables, and known vulnerabilities as a `planning problem.' By applying AI-based planning algorithms, ChainReactor constructs exploitation chains that blend malicious and benign actions to facilitate privilege escalation or lateral movement. This approach reflects a paradigm shift from exploiting isolated vulnerabilities to dynamically engineering multi-step pathways that mirror the complexity and stealth of modern cyberattacks.

\paragraph{Actions on Objectives}
In the last stage, the attacker executes the final objectives, such as data exfiltration which may require evading an outbound firewall. AI may be used to stealthily exfiltrate data through obfuscation, making it look like a different class of data through various AI-techniques.\footnote{\url{https://www.blackhat.com/docs/us-16/materials/us-16-Wolff-Applied-Machine-Learning-For-Data-Exfil-And-Other-Fun-Topics.pdf}}

%% file: Section/05-NeutralTech.tex
\section{Technology is Neutral}
AI, like many technologies, has an inherent dual-use potential: Its impact depends on context and intent of use. While designers envision specific applications, the technology itself remains neutral, and may be used for unintended purposes. For example, DeepExploit was originally designed to automate penetration testing \textit{(defensive)}, but may be re-purposed to exploit vulnerabilities \textit{(offensive)}. Similarly, Adversarial AI can serve for adversarial training to increase the robustness of a classifier, or as means to evade detection. This reflects broader cybersecurity practices, as highlighted in MITRE ATT\&CK software, 
where tools, such as sqlmap \textit{(Initial Access)} or CrackMapExec \textit{(Lateral Movement),} are used by attackers and defenders.

In this regard, an illustrative examples is that of \textbf{Generative Adversarial Networks}. GANs are a form of Generative AI composed of two neural networks: a generator and a discriminator. 
The generator tries to create synthetic data that closely resembles the real training data and attempts to fool the discriminator into classifying this synthetic data as real. In this two-player game, both the generator and the discriminator improve their abilities over time.~\cite{goodfellow2014generative}
GANs are very versatile and, in cybersecurity, can be used in both a benign and malicious way. 
When a GAN is used to generate malware, the interplay of Adversarial AI and Offensive AI \textit{(3)} becomes apparent: MalGAN was designed to generate malware samples that can evade malware detectors \textit{(offensive).~\cite{hu2022generating}} However, the same generated malware samples could also be used to strengthen malware classifiers \textit{ (defensive).}

\vspace{-2pt}
\begin{tcolorbox}[mybox, title={Misconception \#4}]
The following association is incorrect:
\[
\textbf{GAN} = \textbf{Malicious Application of AI} 
\]

Despite having ``adversarial'' in their name, there is \textit{nothing intrinsically malicious} in a GAN. 
The term ``adversarial'' simply refers to the workflow of a GAN, which envisions two neural networks (a generator, and a discriminator) that ``fight'' against each other, with the ultimate purpose of generating new data (which is neutral).

\end{tcolorbox}
\vspace{-2pt}

%% file: Section/06-Interplay.tex
\section{THE INTERPLAY}
The terms `Adversarial AI' and `Offensive AI' -- occasionally replaced by other expressions such as `AI Red Teaming', `AI-driven attacks', or even by writing `ML' instead of `AI' -- are used to describe AI as either \textit{(a)}~an attack surface or \textit{(b)}~an attack vector.
Although both concepts imply distinct threats as described above, there is an overlap among the two, which we pinpoint in the Venn diagram shown in Figure~\ref{fig:interplay}. We identify five distinct \textit{cases} (denoted as `$C\#$' from now on) across the three security domains gravitating around AI and discussed so far in this article: Security of AI Systems; its subset Security of AI models (or ``Adversarial AI''); and Offensive AI. 
In what follows, we provide an organic security-centered discussion on each domain, emphasizing the \textit{cases} in Figure~\ref{fig:interplay}. 

\textbf{Security of AI systems.} As AI models become increasingly adopted and integrated into AI systems, system providers and developers must assess the system-wide vulnerabilities of such AI systems.
Instead of reinventing the wheel, practitioners should first extend established DevSecOps best practices\footnote{DevSecOps best practices refer to the integration of security principles and measures into every phase of the software development life cycle, making security a shared responsibility across development, operations, and security teams.} to AI systems \textit{(C1)}, and then consider the security risks inherent to the AI model. For the former, vulnerabilities in AI-specific libraries, such as TensorFlow (e.g., CVE-2023-33976), should be treated similarly to any other software vulnerability.
This approach aligns with security-by-design principles,~\cite{radanliev2024digital} which mandate that security is a foundational requirement integrated throughout the entire lifecycle of a system---and not an afterthought. For AI systems, this means integrating safeguards against well-known threats such as poisoning and evasion attacks before deploying the overarching system. 
Perhaps intriguingly, AI tools can be used to spot vulnerabilities in AI systems \textit{(C4)}, as demonstrated by solutions like DeepExploit.\footnote{\url{https://github.com/13o-bbr-bbq/machine_learning_security/tree/master/DeepExploit}} DeepExploit leverages AI to autonomously scan, prioritize, and exploit system weaknesses, highlighting AI's dual-use potential for defensive and offensive purposes.

\textbf{Security of AI models.} Nevertheless, given that AI models are intrinsic components of any AI system, DevSecOps practices must evolve to address the unique security risks associated with AI models, specifically in the context of Adversarial AI. Unlike traditional software, which is easier to debug and patch when vulnerabilities are discovered, most AI systems operate as `black boxes,'\footnote{AI models are classified as black box or white box. By default, white box models, such as decision trees, provide transparency in their decision-making process, whereas black box models, like neural networks or LLMs, inhibit such a reasoning due to their complex internal architectures.} making it more difficult to identify, understand and mitigate potential threats. As a result, traditional security measures are insufficient, and new, specialized strategies are required to effectively safeguard against adversarial AI attacks and other model-specific vulnerabilities.
Attacks against AI models can be carried out using basic algorithmic or statistical heuristics \textit{(C2)}, or more sophisticated AI-powered techniques \textit{(C3)}. The attack surface can affect various aspects of the model, including manipulating classifiers, stealing models, or determining if a specific data point was included in the training set.

\textbf{Offensive AI}. Last, AI can be used offensively to violate the security and privacy of any given entity (e.g., people or systems), Yet, such an ``offensive'' use can be multifaceted. For instance, an attacker can use AI for malicious purposes, such as using an LLM to craft spear-phishing emails (C5). However, at the same time, a system defender can use an AI-powered tool to test the security of an AI system (C4) or of an AI model embedded in an AI system (C3). The same cases (C4 and C3) can also be seen in reverse: Attackers can use an AI-powered tool to break an (AI-based) system and cause damage to an organization.

\vspace{-2pt}
\begin{tcolorbox}[mybox, title={Misconception \#5}]
The following association is imprecise:
\[
\textbf{Adversarial AI} = \textbf{Offensive AI} 
\]
Adversarial AI refers to studying the behavior of AI in adversarial environments, whereas Offensive AI refers to using AI to violate the security or privacy of any target. In either case, however, the presence of an ``attacker'' is not a given: Both of these terms include techniques that can be used for good (e.g., making a system more secure) or for bad (e.g., bypassing a system, or deceiving humans). 
\end{tcolorbox}
\vspace{-2pt}

%% file: Section/07-Conclusion.tex
\section{CONCLUSION}
AI is undergoing unprecedented maturation and increasing adoption in real-world applications. LLMs have played a pivotal role in this transformation: Their general-purpose capabilities and intuitive natural language interfaces have driven the widespread adoption among individuals and organizations.
However, AI is not without risks and, as discussed in this article, can be `exploited.' AI can be used to carry out attacks (Offensive AI), but it can also become the target of attacks (Adversarial AI). 
This adds new threats for organizations to consider in modern threat assessments.

Still, we are currently `only'  experiencing the first wave of AI-driven technologies. As such, the associated threat landscape is still in its early stages and continues to evolve. With the ongoing advancement of AI, and particularly LLMs, new paradigms, such as \textbf{Agentic AI}, emerge. An AI agent is an autonomous computational entity that perceives its environment through sensors, makes decisions based on models or policies, and acts upon the environment to achieve specific goals. 
While still hypothetical, these AI agents could, in the near future, support or automate various stages of the cyber kill chain.  Consider that tasks currently performed by human hackers may be delegated to such agents, enabling even more efficient and scalable attacks, specifically if combined with some of the techniques presented in the previous sections, such as Attribute Inference Attacks or smart malware like BlackMamba. 
This possibility underscores the importance of proactively understanding the different types of risks, i.e., Adversarial AI and Offensive AI, as well as their overlaps to not only defend against current AI-driven threats but also against those yet to come.

\vspace*{-8pt}

%% file: Section/bibliography.tex
\vspace*{-8pt}